# THE IMPACT OF BRAZIL ON GLOBAL GRAIN DYNAMICS: A STUDY ON CROSS-MARKET VOLATILITY SPILLOVERS

Felipe Avileis[a] and Mindy L. Mallory[b*]

[a] Graduate Research Assistant, University of Illinois, 1301 W Gregory Dr., Urbana, IL, 61801

[b] Associate Professor and Clearing Corporation Endowed Chair of Food and Agricultural Marketing, Purdue University, 615 West State Street, West Lafayette, IN 47907, USA

* Corresponding author: Tel.: 765-494-4244, email: mlmallor@purdue.edu

# THE IMPACT OF BRAZIL ON GLOBAL GRAIN DYNAMICS: A STUDY ON CROSS-MARKET VOLATILITY SPILLOVERS

**Abstract**

Brazil's rise as a global powerhouse producer of soybeans and corn over the past 15 years has fundamentally changed global markets in these commodities. Brazil's rise is arguably due to the development of varieties of soybean and corn adapted to climates within Brazil, allowing farmers to double-crop corn after soybeans in the same year. Corn and soybean market participants increasingly look to Brazil for fundamental price information, and studies have shown that the two markets have become cointegrated. However little is known about how much volatility from each market spills over to the other. In this article we measure volatility spillover ratios between U.S. and Brazilian first crop corn, second crop corn, and soybeans. We find that linkages between the two countries increased after double cropping corn after soybeans expanded, volatility spillover magnitudes expanded, and the direction of volatility spillovers flipped from U.S. volatility spilling over to Brazil before double cropping, to Brazil spilling over to U.S. after double cropping.

Keywords: corn, soybeans, volatility spillovers, cointegration, GARCH

JEL Codes: Q11, Q13

# THE IMPACT OF BRAZIL ON GLOBAL GRAIN DYNAMICS: A STUDY ON CROSS-MARKET VOLATILITY SPILLOVERS

**Introduction**

Brazil's rise as a global powerhouse producer of soybeans and corn over the past 15 years has fundamentally changed global markets in these commodities. This was possible because the sector went through several structural changes. The biggest being the development of varieties of soybean and corn adapted to climates within Brazil, allowing farmers to double-crop corn after soybeans in the same year. As Brazil's production increased, so did its importance in global grain dynamics. Brazil jumped from a negligible exporter in corn to being the second largest exporter and is now the largest soybean producer and exporter in the world.

While global grain price discovery has historically been located primarily in the Chicago Mercantile Exchange (CME) futures contracts as first described by Garbade and Silber (1983), market participants increasingly are looking to Brazil for fundamental news that will impact global prices (Laca 2017). Yet, to our knowledge, the degree to which information about market fundamentals in U.S. soybean and corn markets are spilling over to Brazilian markets, and vice versa, is unknown. In this paper we estimate volatility spillover ratios between U.S. and Brazil corn and soybean markets before and after the surge in second crop corn. In the case of corn, we conduct analyses with a market from the region where first crop corn is grown (south) and a market where second crop corn is grown (center-west).

Prior work has studied the integration of U.S. and Brazilian markets. Balcombe et al. (2007) analyzed how Brazilian, Argentine and US grain markets were related, investigating possible threshold effects. Utilizing data up to 2006, their study finds that threshold effects exist, and transmission was bigger for corn markets and causality flowed from US and Argentine markets to Brazil. Mattos and Silveira (2015) measure the impacts of the second corn crop in Brazil on seasonality, basis behavior and integration to international markets. They find that after double cropping begins Brazilian markets became more integrated into international markets. Similarly, Li and Hayes (2016) find that U.S. and Brazilian soybeans are integrated with a seasonal effect. Whereby, U.S. futures lead Brazilian futures except for during the Brazilian growing season, when Brazil leads the U.S. soybean price. Our work is most similar to Cruz Jr. et al. (2016) who also utilize futures and cash prices for soybeans and corn. They estimate market integration and test for volatility transmission with the causality in variance test (Cheung and Ng 1996). They find that market integration increased and volatility transmission increased in the corn markets after 2007; in the soybean market they find integration increases as well, but they find that U.S. markets caused volatility in Brazilian markets both before and after 2007.

Despite several studies that have established integration between Brazilian and U.S. soybean and corn prices, to our knowledge none have estimated conditional volatility spillovers between these two important global markets. Cruz Jr. et al (2016) comes the closest to our work. However Cruz et al. use the variance causality test to test for volatility transfer, which is an unconditional test of the null hypothesis of no causality-in-variance. While this approach is well suited to answer the yes-or-no question of whether or not there is volatility transfer, this article is interested in modeling the dynamics in the volatility spillovers.

Managing exposure to volatility is a key part of any physical commodity market participant. As global markets become more integrated, so too do individual markets become more susceptible to market gyrations occurring in other major markets. In this paper we make three contributions to the literature. First, we develop a method for calculating volatility spillover ratios when all markets are endogenous. Volatility spillover ratios are a convenient way to measure how much one market's volatility makes its way into another. This calculation is straightforward when an exogenous market spills over to another that is integrated with the (exogenous) leader (as in Wu, Guan, and Myers, 2011). When all markets are endogenous, calculating a clean transmission of volatility is not straightforward. We propose a method to calculate volatility spillover ratios that isolates volatility transmission to the degree possible, allowing us to produce a visual representation of when and how strong volatility transmission is happening.

Second, we perform our analysis with two different Brazilian corn prices, one from a first crop corn region and one from a second crop corn region. This allows us to determine if markets for the second crop, which is largely destined for export, is more integrated and more likely to give and receive volatility spillovers from U.S. corn prices than the first crop, which is predominantly consumed by domestic livestock.

Third, we document that volatility spillover dynamics in soybeans have been completely inverted since double cropping corn after soybeans took off in 2010. We show that prior to 2010, U.S. soybean price volatility would spillover to Brazil, but not vice versa. Since 2010, Brazil soybean price volatility spills over to the U.S., but U.S. volatility does not spillover to Brazil.

**Background**

The rise of Brazilian production has been fundamental to keeping increasing global demands sated.

Planted Area, Million Hectares

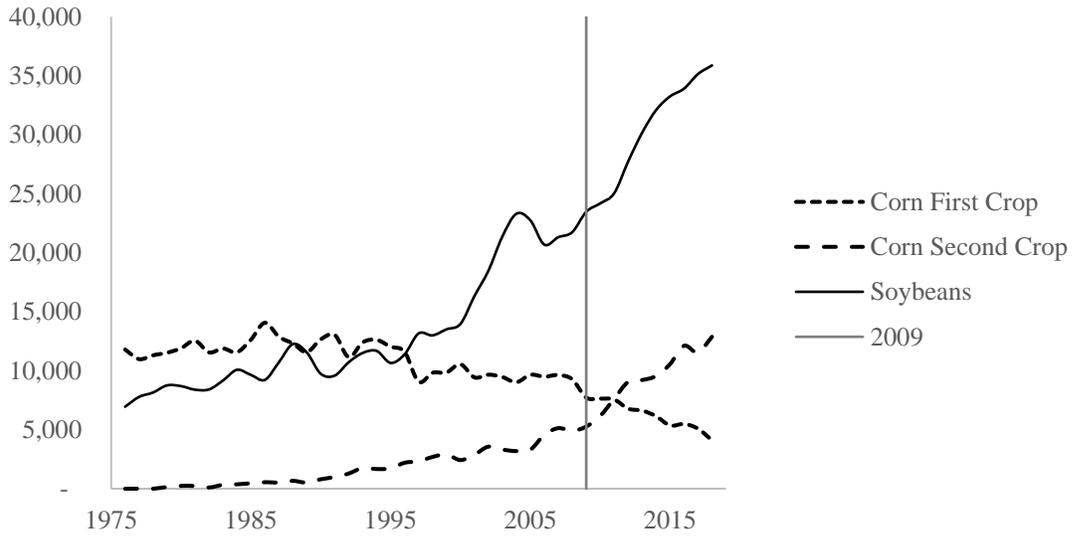

Production, Million Tonnes

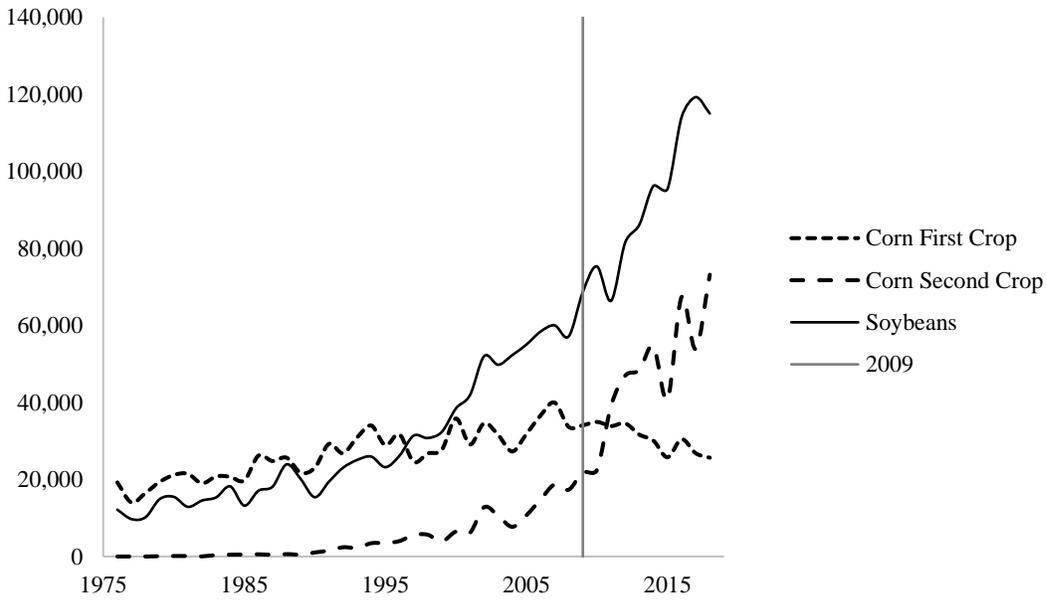

**Figure 1. Brazilian Planted Area and Production of Soybeans, First Crop Corn, and Second Crop Corn, 1976-2019**

Source: CONAB[1]

---

[1] https://www.conab.gov.br/

In figure 1 we display Brazilian area planted and production for soybeans, first crop corn, and second crop corn. In the 1970's researchers discovered that by adding phosphorus and lime to the soil in the country's Cerrado region, the land could be used for commercial agriculture, since the Cerrado receives plentiful rainfall through the growing season.

Around the year 2000 we see that planted area and production of soybeans and second crop corn start to increase rapidly. In the early 2000's researchers and farmers discovered the possibility of double cropping corn after soybeans in the same growing year, effectively doubling the available area for planting (as now farmers could grow both grains on the same space). Additionally, the government's effort to develop corn and soybeans cultivars better adjusted to the region's climate has resulted in dramatic increases in yields in the region, almost doubling over the past 10 years. The development of this region as a big producer of grain is key to understanding the surge of Brazil as a top grain producer and exporter. With an average growing season of 120 days for soybeans and 150 days for corn, farmers planning to double crop plant soybeans as early as possible, usually starting in October, and plant corn right after harvesting beans, as there is little room for wasted days before the dry season arrives at the end of May. This allows the same area to be used for both crops, increasing production for both. Figure 2 illustrates soybeans and corn planting and harvesting months.

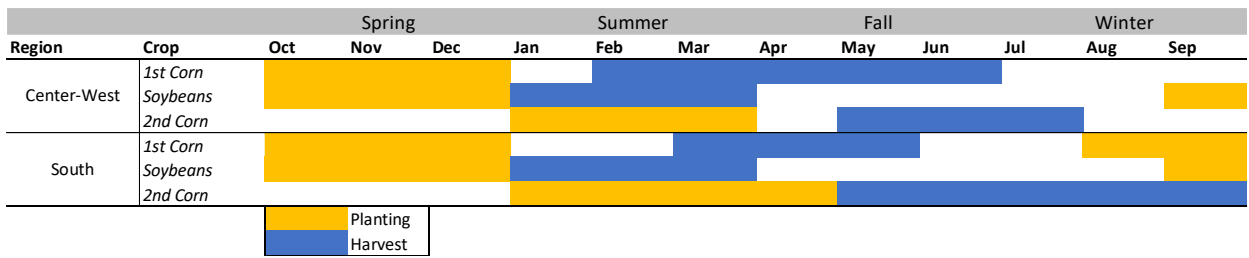

**Figure 2. Crop Calendar**

Source: Conab

One important factor is that if the farmer chooses to plant corn as a first crop, he loses the opportunity to double-crop, as corn takes longer to mature when compared to beans (30 day difference). We posit that this changed the planting decision drastically and has impacted the price responsiveness of Brazilian farmers. Before Brazil was able to double-crop, Brazilian farmers had a similar planting decision as US farmers. Farmers would allocate land to corn versus soybeans and other crops based on which was likely to be the most profitable in that year, given whatever logistical constraints they face related to crop rotations, etc.[2] Land allocation, then, would bring prices to equilibrium as a high corn/soy price ratio would 'buy' acres for corn and a low corn/soy price ratio would 'buy' acres for soybeans. Therefore Brazilian and U.S. farmers had similar levels of pre-planting price responsiveness, and, by default, similar reactions to price shocks.

However, Brazilian farmers who can double crop are not price sensitive, planting both will almost without exception dominate planting one or the other. Considering that the only combination for double-cropping is soybeans then corn, farmers will plant soybeans and then corn, regardless of the price ratio between both. We argue later that this change in price responsiveness of Brazilian farmers has changed volatility transmission in global corn and soybeans markets.

Another dynamic in the Brazilian corn market is that most of the first crop corn is now grown in the southern region, where it is too cold to capitalize on double cropping. This also happens to be the location of the country's large livestock sector, which consumes most of the first corn crop. Meanwhile, much of the second crop corn is exported and uses the same transportation and port infrastructure that is used to export soybeans, just a few months later.

---

[2] We acknowledge that other factors affect a farmers decision on what to plant. However, several studies similar to Miao, Khanna and Huang (2015) point that price is the main driver.

Therefore, we posit that local prices in the first crop corn region will be less connected to international markets than the second crop corn region.

**Data**

In our study we use two Brazilian spot corn prices and one Brazilian spot soybean price along with U.S. futures prices. Our period of study is February 2004 to February 2019. We chose a common beginning period of February 2004 to align the studies, because this is a date for which all series are available, and it is after Brazil had begun double cropping corn. The soybean study could have included a few additional months because data is available and Brazil was already exporting soybeans, however results are qualitatively robust to alternative start dates within this window. The end of the study period was the data available at the time we accessed the data. For U.S. market data we use log of daily Chicago Board of Trade futures closing prices for soybeans ($S$) and corn ($C$). We create a series of nearby prices, rolling an expiring contract forward when the next contract to expire surpasses the expiring contract in volume. Brazilian prices are the daily average price at the Port of Paranaguá, for soybeans, and daily average prices from Chapecó and Sorriso for corn. These prices are from the Center for Advanced Studies on Applied Economics (CEPEA)[3] and are converted to US$/bu using daily foreign exchange rates from the Federal Reserve Economic Database at the St. Louis Federal Reserve Bank of the U.S.[4] Figure 3 illustrates the Brazilian price locations.

The yellow dot shows the location of the port of Paranagua. Data from COMEX STAT shows that the Port of Paranagua exported approximately 15 million tons of soybeans, second only to the Port of Santos (20 million tons).[5] The difference between the two ports is that, while

---

[3] CEPEA is a research center in the Department of Economy, Administration, and Sociology at the University of São Paulo
[4] FRED url https://fred.stlouisfed.org/
[5] COMEX STAT is a database of Brazilian foreign trade maintained by the federal government of Brazil.

Santos only exports grain coming from the Center-West and Southeast, Paranagua exports grains from these regions and also grain originated in the South, so we use this location for Brazilian soybeans prices. The red dot is located at Chapeco. Farmers in this area do not double-crop; they plant first crop corn. This location is also the home of most domestic livestock feed demand because most poultry and pork feedlots are in this area. The green dot is located at Sorriso, in the heart of the biggest second crop corn regions.

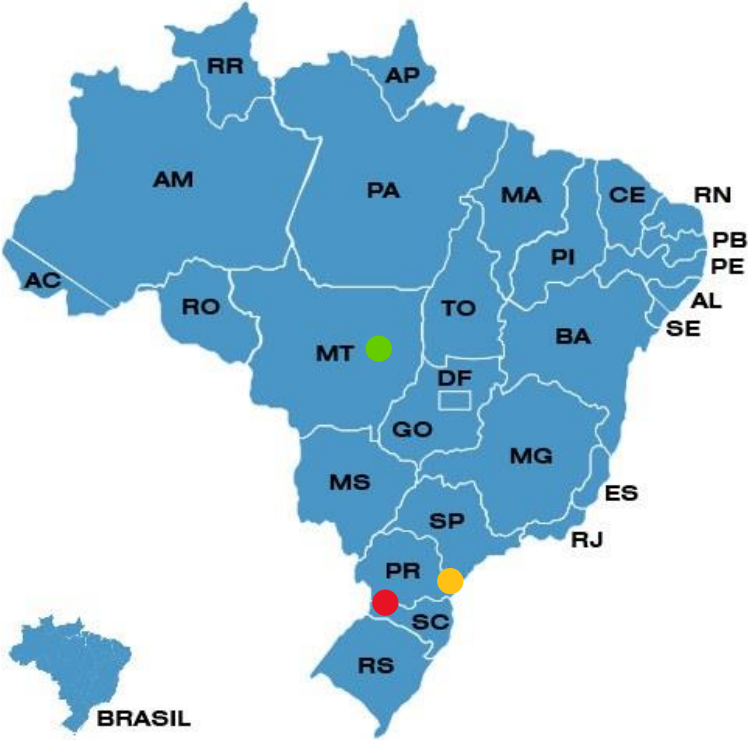

**Figure 3. Cash Prices Location in Brazil**

Source: Departamento Nacional de Infraestrutura de Transportes[6]

Figure 4 shows the log prices across the time studied. The vertical lines indicate where we separate our sample into two periods: Pre- and Post-second crop corn production. U.S. and

---

[6] http://antigo.dnit.gov.br/

Paranagua prices appear highly correlated and at remarkably similar price levels with Brazil price converted to U.S. dollars.

Soybeans

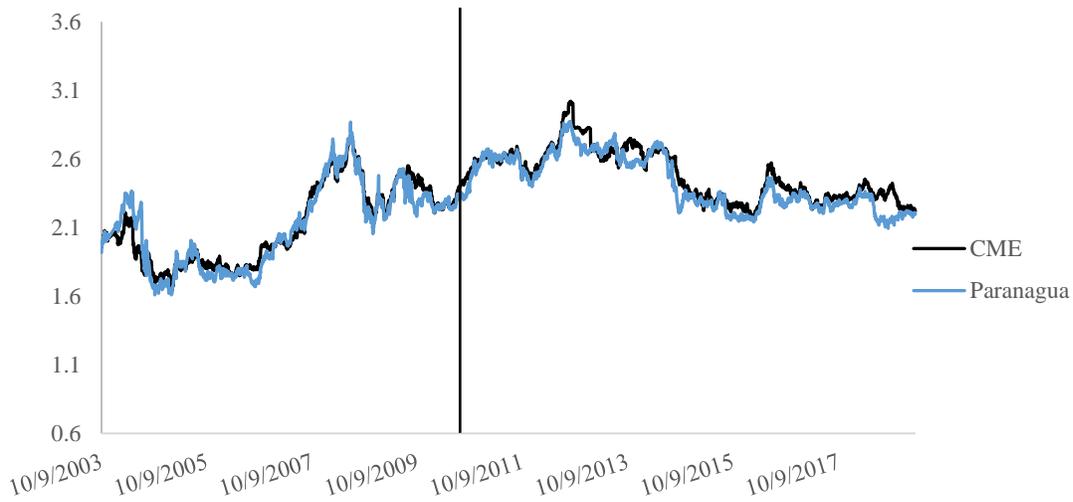

Corn

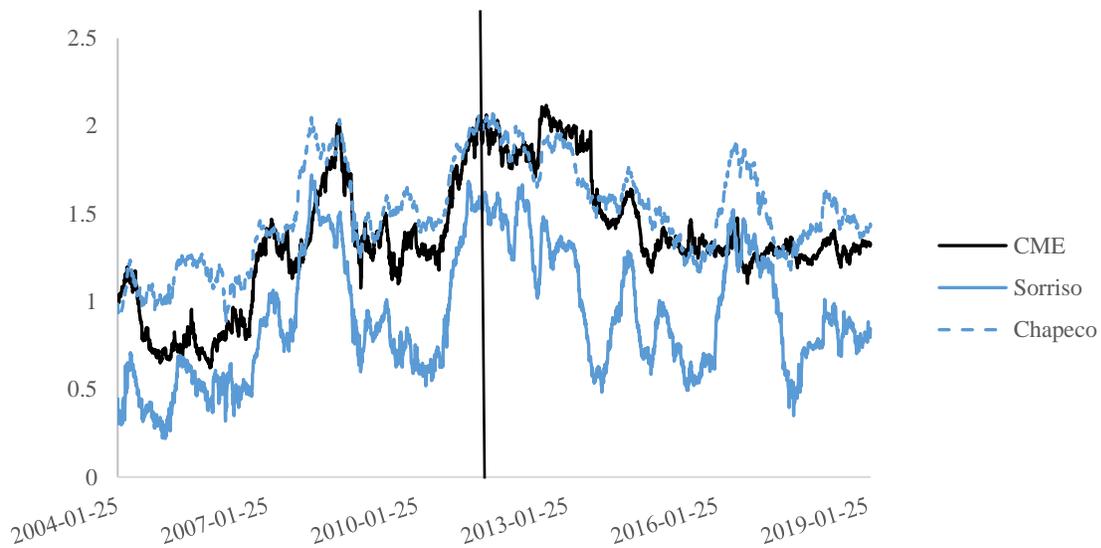

Figure 4: Logarithm of Soybean Prices in U.S. and Brazil, February 1, 2004 to February 1, 2019.

Note: CME is the logarithm of daily Chicago Mercantile Exchange soybean futures prices. Paranagua is the logarithm of daily average of soybean spot prices at the port of Paranagua, Brazil. Vertical line indicates before and double cropping soybeans and corn became possible in parts of Brazil.

The second panel shows corn log prices for our locations of study. U.S. and Chapeco prices appear relatively highly correlated and tend to maintain similar log price levels, however visually the two location are not as tightly integrated as is the case in soybeans. Sorriso, in the heart of the corn double cropping region, trades at a significant discount to CME and Chapeco prices. This is because corn grown in this area is second crop and is almost exclusively destined for the export market, so significant transport cost to end users depresses the price. The Sorriso and Chapeco price movements appear highly correlated, though.

Table 1 contains summary statistics of all the price series we use in our analysis. CME corn prices are somewhat comparable to Chapeco corn prices in levels with mean log prices of $1.51/bu and $1.37/bu respectively. Sorriso log prices typically are discounted compared to Chapco and CME corn prices. All three corn prices are moderately skewed to the right, and are moderately platykurtic. Returns are centered around zero, have very little skewness, but are leptokurtic.

Similarly for both soybean prices. The CME and Paranagua prices have similar log price levels, very little skewness, and are a bit platykurtic. Returns are centered around zero, have little skewness and are leptokurtic.

**Methods**

Our primary objective is to determine the nature of volatility transmission between the U.S. and Brazil before and after 2010. To accomplish this we will model price dynamics by either a VAR in returns or a VECM depending on whether or not the series are cointegrated. Then we obtain residuals from the price dynamics equation and fit a multivariate GARCH model, allowing us to examine volatility spillovers.

Table 1. Summary Statistics

| Location | Mean | Median | Standard Error | Skewness | Excess Kurtosis | N |
|---|---|---|---|---|---|---|
| Corn | | | | | | |
|    Chapeco | 1.51 | 1.49 | 0.00 | 0.05 | -0.87 | 3578 |
|    CME | 1.37 | 1.32 | 0.01 | 0.05 | -0.42 | 3578 |
|    Sorriso | 0.91 | 0.84 | 0.01 | 0.37 | -0.98 | 3578 |
|    Chapeco_Returns | 0.00 | 0.00 | 0.00 | -0.02 | 5.22 | 3578 |
|    CME_Returns | 0.00 | 0.00 | 0.00 | -0.25 | 7.44 | 3578 |
|    Sorriso_Returns | 0.00 | 0.00 | 0.00 | 0.70 | 7.43 | 3578 |
| Soybean | | | | | | |
|    CME | 2.32 | 2.35 | 0.01 | -0.37 | -0.62 | 3492 |
|    Paranagua | 2.28 | 2.29 | 0.00 | -0.40 | -0.52 | 3492 |
|    CME_Returns | 0.00 | 0.00 | 0.00 | -0.62 | 7.81 | 3492 |
|    Paranagua_Returns | 0.00 | 0.00 | 0.00 | -0.87 | 17.03 | 3492 |

Note: Chapeco, Sorriso, and Paranagua are the log of daily spot prices obtained from Bloomberg. CME are the log of daily prices of futures contract prices for corn and soybeans at CMEGroup. Analysis is from February 2004 to February 2019.

*VAR Model Specification*

Vector autoregression (VAR) models captures the linear intertemporal dependencies between two or more variables. In equation (2), each variable has its own equation, affected by $k$ of its own and the other lagged variables.

$$\Delta br_t = a_{1,0} + a_{1,1}\Delta br_{t-1} + a_{1,2}\Delta us_{t-1} + \cdots + a_{1,2k}\Delta br_{t-k} + a_{1,2k+1}\Delta us_{t-k} + e_{1t}$$
$$\Delta us_t = a_{2,0} + a_{2,1}\Delta br_{t-1} + a_{2,2}\Delta us_{t-1} + \cdots + a_{2,2k}\Delta br_{t-k} + a_{2,2k+1}\Delta us_{t-k} + e_{2t}, \qquad (2)$$

where $\Delta br_t$ and $\Delta us_t$ are returns for Brazilian and US prices, respectively, and the $a_{i,j}$'s are regression coefficients. We will use this specification for price dynamics in the cases where prices are non-stationary and not cointegrated.

*VECM Specification*

In cases where the prices are non-stationary and the two series are cointegrated, we will use a vector error correction model (VECM). The error correction term (*ect*) continually pulls the system of equations toward its long run equilibrium, defined by the term $ect = \beta_0 + \beta_1 br_{t-1} + \beta_2 us_{t-1}$. Equation (4) defines the VECM.

$$\Delta br_t = a_{10} + \alpha_1(\beta_0 + \beta_1 br_{t-1} + \beta_2 us_{t-1}) + a_{1,1}\Delta br_{t-1} + a_{1,2}\Delta us_{t-1} + \cdots$$
$$+ a_{1,2k}\Delta br_{t-k} + a_{1,2k+1}\Delta us_{t-k} + e_{1t}$$

$$\Delta us_t = a_{20} + \alpha_2(\beta_0 + \beta_1 br_{t-1} + \beta_2 us_{t-1}) + a_{2,1}\Delta br_{t-1} + a_{2,2}\Delta us_{t-1} + \cdots$$
$$+ a_{2,2k}\Delta br_{t-k} + a_{2,2k+1}\Delta us_{t-k} + e_{2t}, \qquad (4)$$

where $\Delta br_t$ and $\Delta us_t$ and the $a_{i,j}$'s are defined as before and the $\alpha'$s are regression coeficients on the *ect*, and function as speed of adjustment parameters for each series as it is pushed back to long run equilibrium.

In the next step we use residuals from the price dynamics equations, $\varepsilon_t = [e_{1t}, e_{2t}]$, to fit the GARCH model that will help us measure volatility spillovers.

**BEKK GARCH and Volatility Spillover Estimation**

When dealing with more than one variable, a multivariate GARCH (MGARCH) can capture interactions and volatility spillovers. While several different specifications for MGARCH exist, we use the BEKK-GARCH, defined by Engle and Kroner (1995).

Using the BEKK model specification guarantees some important benefits against other MGARCH models, like the VEC-MGARCH in Bollerslev, Engle, and Wooldridge (1988) and the Dynamic Conditional Correlation (DCC) specifications (Bollerslev 1990; Engle 2002). The BEKK, which restricts a VEC-type model to ensure the conditional covariance matrix is positive definite. Further we choose the BEKK-GARCH over the DCC-GARCH because the conditional correlations are flexible in the BEKK model, while the DCC requires them to be static.

Equation 5 shows the BEKK-GARCH representation that guarantees that $H_t$ (the conditional variance-covariance matrix) is positive semi-definite;

$$H_t = C'C + \sum_{j=1}^{q}\sum_{k=1}^{K} A'_{kj} e_{t-j} e'_{t-j} A_{kj} + \sum_{j=1}^{q}\sum_{k=1}^{K} B_{kj}' H_{t-j} B_{kj}, \qquad (5)$$

where $C$, is a lower triangular matrix, $A_{k,j}$, $B_{k,j}$ are N x N parameter matrices, and $e_{t-j}$ are residuals from the mean equation system. Considering a two-variable system (*br* and *us*), equation can be written as in equation 6 for additional clarity.

$$\begin{bmatrix} h_{brbr,t} & h_{brus,t} \\ h_{brus,t} & h_{usus,t} \end{bmatrix} = \begin{bmatrix} c_{11} & 0 \\ c_{21} & c_{22} \end{bmatrix} \begin{bmatrix} c_{11} & 0 \\ c_{21} & c_{22} \end{bmatrix}' + \begin{bmatrix} a_{11} & a_{12} \\ a_{21} & a_{22} \end{bmatrix}' \begin{bmatrix} e_{br,t-1}^2 & e_{br,t-1}e_{us,t} \\ e_{br,t}e_{us,t-1} & e_{us,t-1}^2 \end{bmatrix}$$
$$\begin{bmatrix} a_{11} & a_{12} \\ a_{21} & a_{22} \end{bmatrix} + \begin{bmatrix} b_{11} & b_{12} \\ b_{21} & b_{22} \end{bmatrix}' \begin{bmatrix} h_{brbr,t-1} & h_{brus,t-1} \\ h_{brus,t-1} & h_{usus,t-1} \end{bmatrix} \begin{bmatrix} b_{11} & b_{12} \\ b_{21} & b_{22} \end{bmatrix}. \qquad (6)$$

This defines the conditional volatilities of *br* and *us* to be as in equation 7.

$$h_{brbr,t} = c_{11}^2 + a_{11}^2 e_{br,t-1}^2 + 2a_{11}a_{21} e_{br,t-1} e_{us,t-1} + a_{21}^2 e_{us,t-1}^2 + b_{11}^2 h_{brbr,t-1}$$
$$+ 2b_{11}b_{21} h_{brus,t-1} + b_{21}^2 h_{usus,t-1}$$

$$h_{usus,t} = c_{12}^2 + c_{22}^2 + a_{12}^2 e_{br,t-1}^2 + 2a_{12}a_{22} e_{br,t-1} e_{us,t-1} + a_{22}^2 e_{us,t-1}^2 + b_{12}^2 h_{brbr,t-1}$$
$$+ 2b_{12}b_{22} h_{brus,t-1} + b_{22}^2 h_{usus,t-1} \qquad (7)$$

*Defining Volatility Spillovers*

In our context where the markets both influence one another, volatility spillovers are complicated to tease out compared to situations where volatility in an exogenous market spills over to other markets (Trujillo, Mallory, and Garcia 2012). In the case of endogenous markets, we propose to use the all the terms with cross market conditional variances and conditional covariances, as well as the square of the cross market residual term from (7) to construct a measure of volatility spillover. The spillover ratio from U.S. to Brazil is shown in equation (8) and the volatility spillover ratio from Brazil to U.S. is shown in equation (9).

$$SR_{us \rightarrow br} = \frac{(a_{21}^2 e_{us,t-1}^2 + b_{21}^2 h_{usus,t-1} + 2b_{11}b_{21}h_{brus,t-1})}{h_{brbr}} \quad (8)$$

$$SR_{br \rightarrow us} = \frac{(a_{12}^2 e_{br,t-1}^2 + b_{12}^2 h_{brbr,t-1} + 2b_{12}b_{22}h_{brus,t-1})}{h_{usus}} \quad (9)$$

As the BEKK defines conditional volatility as always positive and the fact that price shock effects are squared, the ratio automatically assumes a positive or zero value (unless a negative covariance coefficient is present and significant). Also notice that the SR can be higher than 100%. Although the cross-market coefficients will always be positive, by definition, the other coefficients in the calculation are not necessarily positive. The interaction coefficients (i.e. $2a_{11}a_{21}e_{br,t-1}e_{us,t-1}$) from (8) and (9) can assume negative values (for example when a shock is negative and the coefficient is positive, or vice versa).

This definition derived from the BEKK conditional volatilities equation will allow the calculated Spillover Ratios to be bigger than 100%. There are two interpretations of this phenomenon: 1) Volatility at *t* is markedly lower than at *t-1*, so the volatility and price shocks effects from *t-1* are representing most of the volatility at *t*; 2) The Spillover Ratio presents the highest effect one market can have on the other. However, that effect also counts for the covariance effect, which can also count for endogenous effects from the target market on itself.

Although not a perfect measure, this approach to calculating the cross-market effects, or spillovers, allow this study to calculate varying *t*-to-*t+1* ratios. This is key to evaluate seasonal effects between markets and the evolution of the relationships, a key feature of this study. This has an advantage over other methods, like calculating the impulse response function, used by Gardebroek and Hernandez (2013) because it makes use of the changing variance-covariance matrix to show the variation in how markets are spilling over to one another through time. An impulse response function requires a fixed variance-covariance matrix, which would simply average-out the BEKK estimations. It is also is advantageous over the regression method used in Wu, Guan and Myers (2011) when there is not one clear exogenous market spilling over to other markets(s) because it does not require the series to contain an exogenous market.

**Results**

All the prices in this analysis are I(1), non-stationary in levels and stationary in returns.[7] Table 2 shows the results of running pairwise cointegration analysis using Johansen (1991). Prior to 2010 none of the U.S. and Brazilian markets are cointegrated. After 2010 all price pairs are cointegrated. Based on the cointegration results we fit all U.S.-Brazil log price pairs from 2004-2009 to a vector autoregression in levels; we fit all U.S.-Brazil log price pairs from 2010-2019 to a vector error correction model.

*VAR and VECM Results*

Table 3 shows the cointegrating vectors for each of the price pairs. We see the discount on Sorriso corn reflected in the constant value of 0.68 in the CME/Sorriso cointegrating relationship that reflects the increased transport cost of getting the second crop corn to export markets. The other beta coeficients, at 1 and -1.07 for Brazil and CME respectively show that aside from this constant wedge, the long run relationship for CME and Sorriso soybean log prices is to move fairly closely in a one-for-one fashion.

Table 2: Cointegration Analysis

---
[7] See supplement for stationarity testing results.

|  | Pre | Post | 10pct | 5pct | 1pct |
|---|---|---|---|---|---|
| Corn - CME/Sorriso | | | | | |
| r <= 1 \| | 2.02 | 2.14 | 7.52 | 9.24 | 12.97 |
| r = 0 \| | 11.45 | 22.65 | 17.85 | 19.96 | 24.60 |
| Corn - CME/Chapaco | | | | | |
| r <= 1 \| | 2.57 | 1.25 | 7.52 | 9.24 | 12.97 |
| r = 0 \| | 10.13 | 26.32 | 13.75 | 15.67 | 20.20 |
| Soybean - CME/Paranagua | | | | | |
| r <= 1 \| | 1.40 | 1.86 | 7.52 | 9.24 | 12.97 |
| r = 0 \| | 12.37 | 29.87 | 13.75 | 15.67 | 20.20 |

Note: Johansen (1991) test of cointegration between CME and Brazilian price pairs. Pre is 2003-2009 and Post is 2010-2019. The columns 10pct, 5pct, and 1pct show the 10 percent, 5 percent, and 1 percent critical values of the test statistic for the null hypothesis stated by rows.

CME and Chapeco on the other hand are not as tightly bound. With a negative constant and a -0.85 for the beta coefficient on the CME log price we see that in the long run when CME corn log prices are low, Chapeco corn will tend to be higher than CME, but for high CME log prices, Chapeco corn log prices will be below CME.[8] For CME and Paranagua log price we see the long run relationship is tight with the constant equal to 0.12 and with a beta coefficient on the CME term of -1.03.

Table 3: Long-Run Equilibrium Relationships from VECM Models

|  | **CME/Sorriso** | **CME/Chapaco** | **CME/Paranagua** |
|---|---|---|---|
| Brazil | 1 | 1 | 1 |
| CME | -1.07 | -0.85 | -1.03 |
| Constant | 0.68 | -0.26 | 0.12 |

Note: Columns show the beta vector from the VECM model for the 2010-2019 periods for each U.S.-Brazil price pair.

---

[8] That is, we can see this graphing the line defined by 0 = Chapeco – 0.85CME – 0.26 and comparing to the line y = x.

In table 4 we have the VAR and the rest of the VECM model results. The table is organized so that results for equations are presented in rows and the price pairs that are estimated together are separated by the dotted barriers between rows. Pre and Post subperiod model results are presented within these groups. Only Post subperiods have coefficient estimates for an error correction term (ect) included in the model because in the Pre subperiod a VAR model was used. Lag length for all specifications was chosen by Bayesian Information Criterion. In all cases a lag length of two was chosen except for CME and Paranagua in the Post period; in which case, three lags were chosen. We include a dummy variable for 2016 since there was a severe drought in Brazil that especially impacted the corn market, particularly in how much second crop was available for export.

Also in table 4 we show that in the case of corn, prior to 2010 there was little connection between U.S. and the Brazilian corn markets at Sorriso and Chapeco. In both cases the first lag of the CME price is the only significant coefficient and in both cases it is only significant at the 10% level. Likewise there are no significant Brazil terms in the CME corn equations. So neither Sorriso nor Chapeco corn prices had an impact on CME corn prices prior to 2010.

In the case of soybeans before 2010 we find significant connections between Paranagua and CME log prices, even though we did not find cointegration. The coefficient on the first CME lag is 0.1724 indicating the two prices are positively related and the relationship is highly significant (at the 0.001 level). The second CME lag in the Paranagua equation is similar is size at 0.1338 and significant at the 0.01 level. Likewise, in the CME price equation the coefficient on the first lag of the Paranagua price is 0.1177 and significant at the 0.001 level.

In the Post 2010 subperiod we see all three markets demonstrate linkages. All three price pairs are cointegrated. In the case of corn CME short run effects are significant in the Brazilian price equations, but the Brazilian prices are only influencing the CME price through the long run dynamics. Between Sorriso and CME, the speed of adjustment to long run equilibrium (measured by the alpha's on the ECT) are of similar size, indicating that neither market is the clear leader in the long run dynamics. Between Chapeco and CME a different story emerges. The size of the speed of adjustment parameter on Chapeco is about 60% larger than the speed of adjustment parameter on CME corn, indicating that CME corn is the price leader and the Chapeco price responds more strongly to restore the equilibrium. Given that Sorriso is the home of the Safrina crop which competes in the export market with CME corn, while Chapeco corn is used for domestic livestock feeding the stronger link between CME and Sorriso is expected.

For soybeans in the Post period we see significant long run and short run dynamics between Paranagua and CME soybean prices. The cross market short run effects are highly significant in both equations. The long run relationship is somewhat surprising. The alpha on the ECT is significant in the CME equation but not in the Paranagua equation, indicating that Paranagua prices are independent of, or leading, the market and U.S. prices do most of the work adjusting to maintain the equilibrium.

*BEKK-GARCH Results*

In table 5 we show the results of taking the residuals from the VAR and VECM models and fitting them to a bivariate BEKK model to capture any volatility spillover dynamics that may be at play. This table is the result of estimating the BEKK-GARCH model shown in equation 6 (including the mu parameters which centers the residuals around zero). We see the constant and own equation effects are highly significant in nearly all cases ($c_{11}$, $c_{22}$, $a_{11}$, $a_{22}$, $b_{11}$, $b_{22}$), as we would expect. To determine if there are significant volatility spillovers we are most interested in the cross market coefficients we have highlighted with the dotted boxes ($a_{21}$, $a_{12}$, $b_{21}$, $b_{22}$).

Table 4: VAR and VECM Results

|  |  | ECT | BR-1 | CME-1 | BR-2 | CME-2 | BR-3 | CME-3 | 2016 Dummy |
|---|---|---|---|---|---|---|---|---|---|
| Pre | Sorriso | – | -0.1650*** (0.0271) | 0.0877* (0.0352) | -0.0565* (0.0271) | 0.0284 (0.0352) | – | – | – |
|  | CME | – | 0.0095 (0.0209) | 0.0278 (0.0271) | 0.0127 (0.0209) | 0.0133 (0.0271) | – | – | – |
| Post | Sorriso | -0.0079** (0.0027) | -0.0839*** (0.0216) | 0.1534*** (0.0301) | -0.0275 (0.0215) | 0.0606* (0.0303) | – | – | 0.0073*** (0.0022) |
|  | CME | 0.0065*** (0.0019) | 0.0258. (0.0155) | 0.0186 (0.0216) | -0.0150 (0.0154) | -0.0512* (0.0217) | – | – | -0.0038* (0.0016) |
| Pre | Chapeco | – | -0.1144*** (0.0272) | 0.0497* (0.0217) | 0.0274 (0.0272) | 0.0312 (0.0217) | – | – | – |
|  | CME | – | 0.0076 (0.0341) | 0.0290 (0.0272) | -0.0209 (0.0341) | 0.0191 (0.0272) | – | – | – |
| Post | Chapeco | -0.0118*** (0.0029) | -0.1088*** (0.0214) | 0.0755*** (0.0189) | -0.0150 (0.0214) | 0.0448* (0.0190) | – | – | 0.0056*** (0.0014) |
|  | CME | 0.0075* (0.0033) | -0.0047 (0.0245) | 0.0253 (0.0215) | 0.0275 (0.0244) | -0.0490* (0.0216) | – | – | -0.0028. (0.0016) |
| Pre | Paranagua | – | -0.1238*** (0.0328) | 0.1724*** (0.0474) | -0.1370*** (0.0335) | 0.1338** (0.0479) | -0.0726* (0.0331) | 0.0887. (0.0467) | – |
|  | CME | – | 0.1177*** (0.0228) | -0.1026** (0.0329) | 0.0320 (0.0233) | -0.0005 (0.0333) | -0.0025 (0.0230) | .0000 (0.0324) | – |
| Post | Paranagua | -0.0059 (0.0051) | -0.0884*** (0.0235) | 0.0720** (0.0258) | -0.0017 (0.0236) | 0.0149 (0.0256) | – | – | 0.0005 (0.0010) |
|  | CME | 0.0188*** (0.0046) | 0.1088*** (0.0211) | -0.0465* (0.0231) | 0.0865*** (0.0211) | -0.0269 (0.0229) | – | – | 0.0008 (0.0009) |

Note: Pre models are VAR models estimated on 2004-2009 data. Post models are VECM models estimated on 2010-2019 data. Lag length selected with Bayesian Information Criterion. Brazil experienced a severe drought in 2016. A dummy variable was added to capture how this event affected the relationship between these markets. Significance indicators: '***', $p < 0.001$; '**', $0.001 < p < 0.01$; '*'. $0.01 < p < 0.05$; '.', $0.05 < p < 0.10$.

In the case of corn we do not see evidence of strong spillovers. In the Pre period none of the cross market coefficients are significant. In the Post period we do see the a21 and a12 cross market coefficients become significant for Sorriso/CME and a21 become significant for Chapeco/CME. So there are modest spillovers back and forth between CME and Sorriso and from Chapeco to CME in the Post period, but the magnitude of these spillovers is not a large proportion of total volatility as we will see when we plot the spillover ratios in a later section.

As was the case for our mean equation analysis, the results for soybeans tell a different story. Already in the Pre period we see significant cross market effects between Paranagua and CME. The nature of the volatility spillovers is markedly different in the Pre and Post samples, however. In the Pre sample, the only significant terms are a12 and b12, the coefficients that govern the spillover *from* CME *to* Paranagua. In the Post sample the only significant terms are the a21 and b21 coefficients, the coefficients that govern the spillover *from* Paranagua *to* CME. So in the Pre sample we see volatility spillovers from CME to Paranagua, but not from Paranagua to CME, while in the Post sample the direction of spillovers are reversed. Volatility spillovers go from Paranagua to CME, but not the other way. Further, the magnitude of the soybean spillovers is about 100 times what we estimated the corn spillovers to be. Plots of the conditional variances and covariances implied by these estimates are found in the appendix.

*Visualizing the Volatility Spillover Ratios*

We defined volatility spillover ratios in equations 8 and 9. These take the cross market contributions to volatility and divides that by the (total) conditional volatility for each time period. This allows us to visualize volatility spillover ratios similarly to how we can plot the path of conditional volatility and covariance.

Table 5: BEKK GARCH Model on VAR/VECM Residuals

|     | Sorriso/ CME - Pre | Sorriso/ CME - Post | Chapeco/ CME - Pre | Chapeco/ CME - Post | Paranagua / CME – Pre | Paranagua / CME - Post |
|---|---|---|---|---|---|---|
| mu1 | 0.0001 | -0.0004 | 0.0009** | -0.0001 | 0.0003 | -0.0001 |
|     | (0.00) | (0.00) | (0.00) | (0.00) | (0.00) | (0.00) |
| mu2 | 0.0005 | -0.0004 | 0.0007 | -0.0003 | 0.0006 | -0.0002 |
|     | (0.00) | (0.00) | (0.00) | (0.00) | (0.00) | (0.00) |
| c11 | 0.0039*** | 0.0061*** | 0.0038*** | 0.0040 | 0.0058*** | 0.0027*** |
|     | (0.00) | (0.00) | (0.00) | (0.00) | (0.00) | (0.00) |
| c21 | -0.0010 | -0.0001 | -0.0004 | -0.0001 | 0.0001 | 0.0097*** |
|     | (0.00) | (0.00) | (0.00) | (0.00) | (0.00) | (0.00) |
| c22 | -0.0022*** | 0.0013** | 0.0023*** | 0.0019*** | -0.0014*** | 0.0003 |
|     | (0.00) | (0.00) | (0.00) | (0.00) | (0.00) | (0.01) |
| a11 | 0.2928*** | 0.1865*** | 0.3171*** | 0.1858** | 0.5698*** | 0.1878*** |
|     | (0.02) | (0.03) | (0.04) | (0.08) | (0.05) | (0.05) |
| a21 | 0.0045 | -0.0558*** | 0.0170 | -0.0595*** | 0.0307 | -0.4899*** |
|     | (0.02) | (0.02) | (0.03) | (0.02) | (0.02) | (0.04) |
| a12 | 0.0111 | 0.0694** | -0.0055 | 0.0313 | -0.3520*** | -0.0269 |
|     | (0.0288) | (0.0283) | (0.0237) | (0.0181) | (0.0524) | (0.0284) |
| a22 | 0.1865*** | 0.2409*** | 0.1731*** | 0.2627*** | 0.1920*** | 0.3664*** |
|     | (0.0256) | (0.0222) | (0.0250) | (0.0389) | (0.0280) | (0.0460) |
| b11 | 0.9471*** | 0.9487*** | 0.9204*** | 0.9458*** | 0.8224*** | 0.9683*** |
|     | (0.0076) | (0.0139) | (0.0270) | (0.0608) | (0.0320) | (0.0121) |
| b21 | -0.0009 | 0.0102 | -0.0054 | 0.0138 | 0.0004 | 0.1213*** |
|     | (0.0062) | (0.0077) | (0.0167) | (0.0122) | (0.0102) | (0.0268) |
| b12 | 0.0111 | -0.0041 | 0.0093 | -0.0034 | 0.1246*** | -0.0016 |
|     | (0.0090) | (0.0075) | (0.0092) | (0.0072) | (0.0320) | (0.0461) |
| b22 | 0.9764*** | 0.9677*** | 0.9792*** | 0.9609*** | 0.9721*** | 0.4500*** |
|     | (0.0078) | (0.0056) | (0.0076) | (0.0116) | (0.0098) | (0.0586) |

Note: Estimates of the bi-variate BEKK GARCH models on the residuals of the pairwise VAR/VECM models. Pre models are VAR models estimated on 2004-2009 data. Post models are VECM models estimated on 2010-2019 data. Highlighted boxes show how the interaction terms change in the Pre and Post samples. Significance indicators: '***', $p < 0.001$; '**', $0.001 < p < 0.01$; '*'. $0.01 < p < 0.05$; '.', $0.05 < p < 0.10$.

Figure 5 displays the volatility spillovers from the CME prices to Sorriso and Chapeco, respectively. Both the Pre and Post periods are plotted on one figure so a sharp change is visible between 2009 and 2010 where we break our sample. In the Pre sample period the spillover ratios are zero because we did not have any of the cross market spillover coefficients significant. Spillovers go both ways beginning in 2010 when Brazil's second crop began accelerating.

Comparing the spillovers between CME and Sorriso and Chapeco the spillovers between CME and Sorriso are larger than the spillovers between CME and Chapeco, another result of the second crops greater integration to global markets than the first crop corn dominating Chapeco's market.

In both cases of the CME prices spilling over to Sorriso an Chapeco a seasonal pattern is detectable during middle of the year when weather concerns for the U.S. corn crop tend to be at its highest. Going the other way, we tend to see a spike in volatility spillover coming from the Brazilian markets to CME just after the first of the year until about April when weather concerns are a factor in both the first and second crop corn markets.

The volatility spillovers between CME and Paranagua soybeans are shown in figure 6. First, notice how much more pronounced the volatility spillovers are in the case of soybeans (and note that the y-axis in the soybean figure goes ten times higher than the figures for corn volatility spillovers). Second, this figure shows the dramatic switch in the direction of volatility spillovers that we saw from the BEKK-GARCH results. CME soybean prices were spilling over the Paranagua prices prior to 2010. After 2010 the direction of spillover switches and it also seems like the magnitude of volatility spillovers now from Paranagua to CME is larger than the spillovers from CME to Paranagua were prior to 2010.

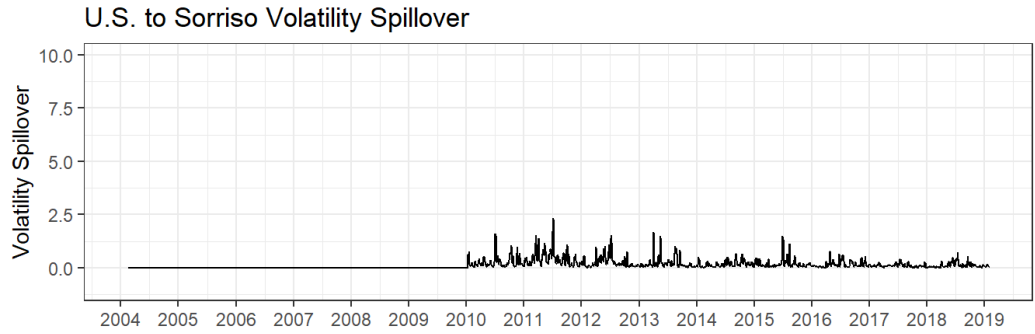
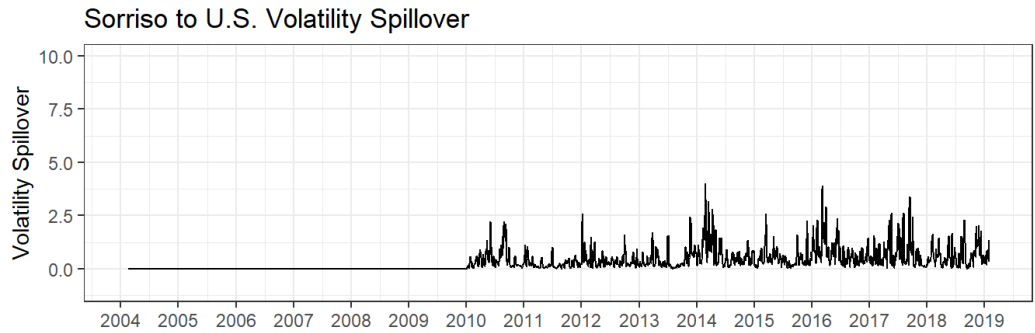
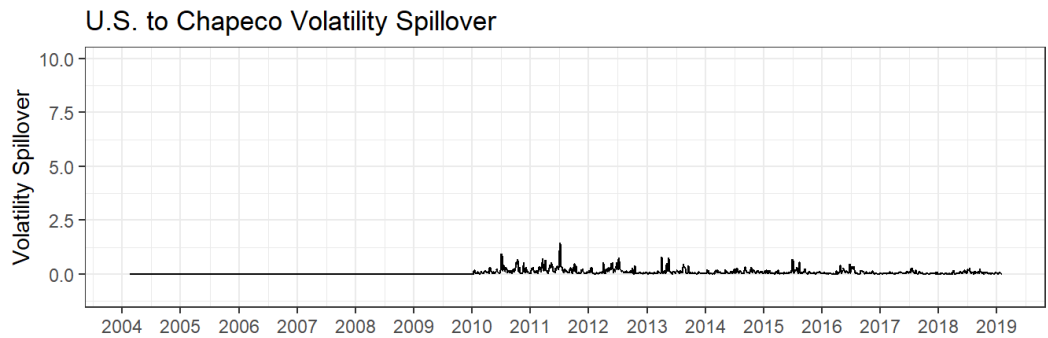
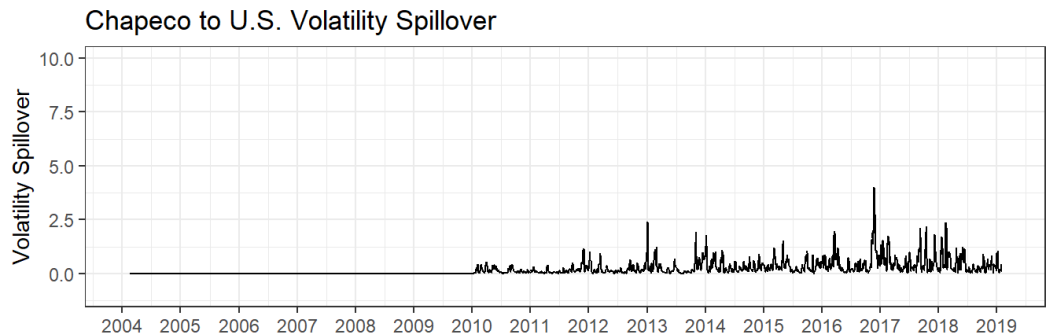

**Figure 5: Corn Volatility Spillover Ratios**

Our result in the soybean market contrasts with the work closest to ours on the topic from Cruz Jr. (2016). They found causality in variance from U.S. to Brazil, but not the other way around. There are three most probable explanations, in our opinion. First, the ability to double crop makes farmers very price insensitive, since double cropping strictly dominates planting first crop corn over a wide range of price scenarios. This price insensitivity may make Brazilian markets more volatile during planting periods, while U.S. farmers' tendency to switch acres between corn and soybeans as expected profits changes dampens price volatility.

Second, Cruz Jr. use average spot prices from the Center-West region in Brazil and the Central Illinois region in U.S., where as we use the single most important market in each case, Paranagua in Brazil, and U.S. futures. We use single market spot prices because using a region average price can aggregate away important variance; given that our primary interest is the study of volatility spillovers, we did not want aggregation to dampen volatility impacts. Third, we use slightly different sample periods, which can affect statistical outcomes.

**Conclusions**

In this article we calculate volatility spillovers between U.S. corn and soybeans prices and Brazilian first crop corn, second crop corn, and soybean prices. We find that prior to the rapid expansion in 2010 of corn after soybean double cropping in the Cerrado region, there were no volatility spillovers between the U.S. and Brazilian corn. U.S. soybean price volatility spilled over to Brazilian soybean price volatility, but not vice versa. After the expansion of corn after soybean double cropping rotations in 2010, all markets examined became cointegrated and volatility began to spillover back and forth between U.S. and Brazil corn markets, with stronger spillovers between Sorriso (the corn spot price from the double cropping region) and the U.S than between Chapeco (the corn spot price from the first crop corn region) and the U.S.

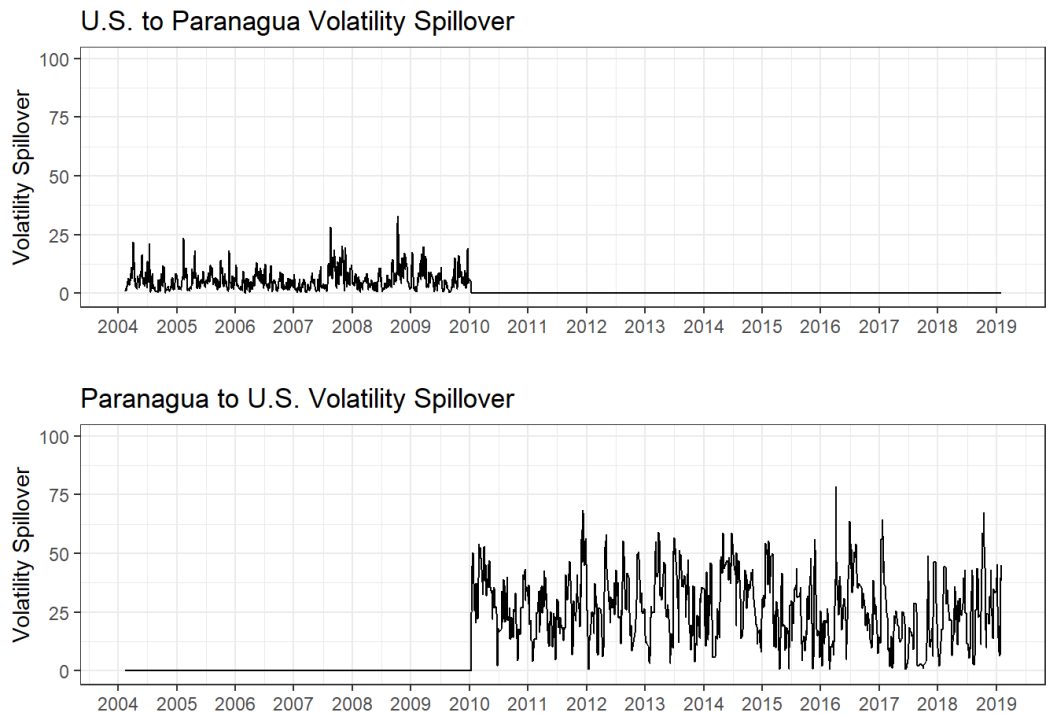

**Figure 6: Soybean Volatility Spillover Ratio**

Additionally, after the expansion of corn after soybean double cropping, the direction of volatility spillovers exactly reversed; there were volatility spillovers from Brazil to U.S., but not the other direction.

Our work documents the growing importance of Brazil and consequently the slightly diminished importance of the U.S. in soybean price leadership. Our work taken together with the recent work of Janzen and Adjemian (2017) suggests that the U.S.' dominant position as the world commodity price leader in soybean and wheat markets may be waning. The long term effects of this are unclear, and should be the topic of further research.

**Appendix**

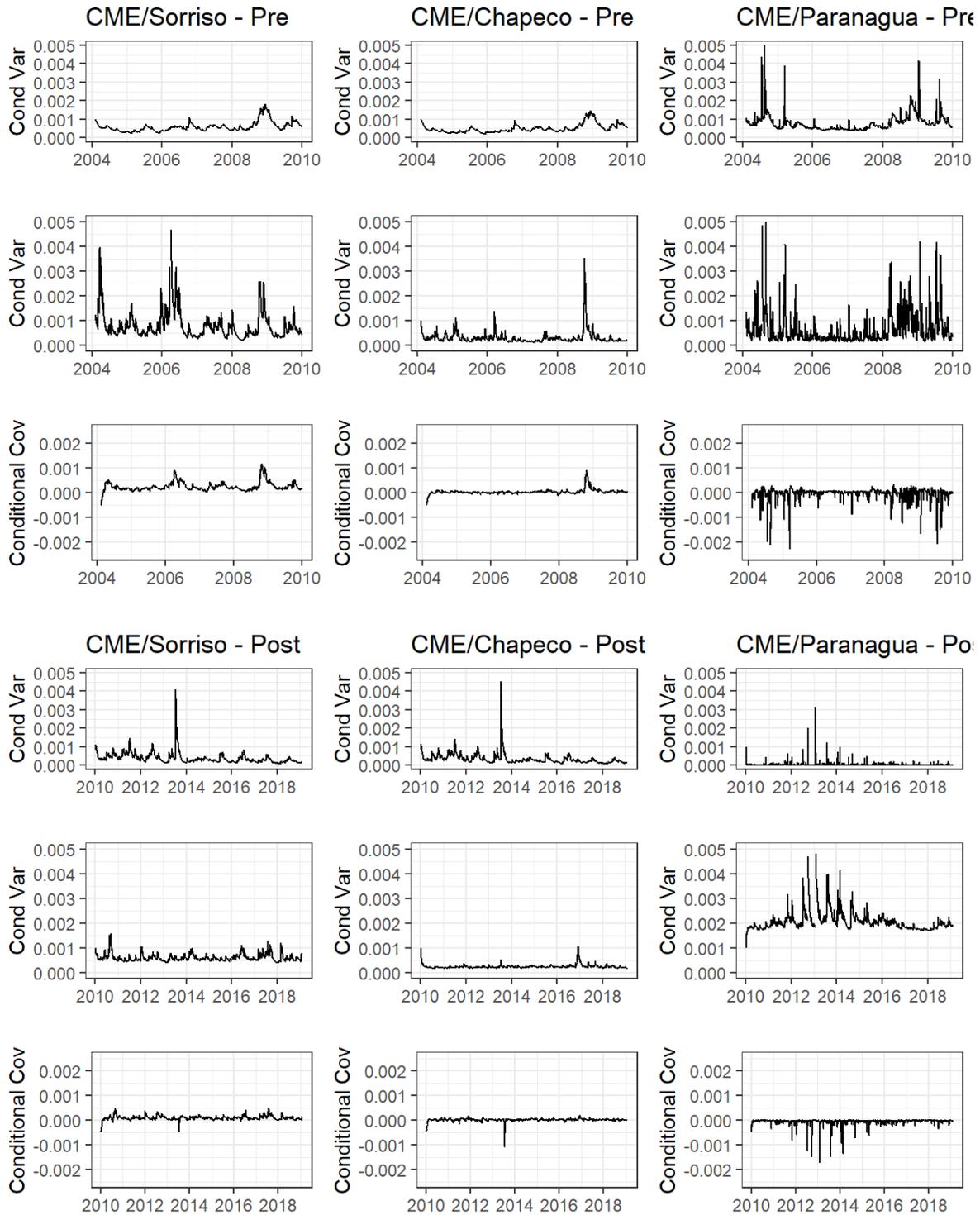

**Figure A1: Conditional Variances and Conditional Covariances for all U.S. Brazil Log Price Pairs**